\documentclass[aps,pra,twocolumn,amsfonts, amssymb, amsmath, amsthm,groupedaddress]{revtex4-1}
\usepackage{url}
\usepackage{color}
\usepackage{graphicx}
\usepackage{amsthm}

\DeclareMathOperator{\Tr}{\operatorname{Tr}}
\usepackage{relsize}
\begin{document}
%\preprint{}

\title{Decomposing qubit positive-operator valued measurements into continuous destructive weak measurements}

\author{Yi-Hsiang Chen$^1$}
\author{Todd A. Brun$^{1,2}$}
\affiliation{$^1$Department of Physics, University of Southern California, Los Angeles, California 90089, USA}
\affiliation{$^2$Communication Sciences Institute, University of Southern California, Los Angeles, California 90089, USA}

\date{\today}

\begin{abstract}
It has been shown that any generalized measurement can be decomposed into a sequence of weak measurements corresponding to a stochastic process. However, the weak measurements may require almost arbitrary unitaries, which are unlikely to be realized by any real measurement device. Furthermore, many measurement processes are destructive, like photon counting procedures that terminate once all photons are consumed. One cannot expect to have full control of the evolution of a state under such destructive measurements, and the possible unitaries allow only a limited set of weak measurements. In this paper, we consider a qubit model of destructive weak measurements, which is a toy version of an optical cavity, in which the state of an electromagnetic field mode inside the cavity leaks out and is measured destructively while the vacuum state $|0\rangle$ leaks in to the cavity. At long times, the state of the qubit inevitably evolves to be $|0\rangle$, and the only available control is the choice of measurement on the external ancilla system. Surprisingly, this very limited model can still perform an arbitrary projective measurement on the qubit in any basis, where the probability of getting an outcome satisfies the usual Born rule. Combining this method with probabilistic post processing, the result can be extended to any generalized measurement with commuting POVM elements. This implies, among other results, that any two-outcome POVM on a qubit can be decomposed into a sequence of destructive weak measurements by this restricted measurement device. 
\end{abstract}

%\keywords{}

\maketitle

\section{Introduction}\label{intro}

A weak measurement is a generalized quantum measurement that disturbs the state of the measured system by only a small amount, but also yields only a very small amount of information about the system, on average \cite{Aharonov88}.  A sequence of weak measurements can be used to describe a continuous measurement process.  The state of the system evolves stochastically, according to a quantum trajectory equation \cite{Brun02}.  While each weak measurement provides only a small amount of information, a sequence of measurements can allow information to accumulate, so that the system undergoes a strong measurement.

Earlier work has shown that any generalized measurement can be decomposed into a series of weak measurements. For a two-outcome measurement, the sequence of weak measurements takes the form of a random walk in one dimension \cite{Oreshkov05,Oreshkov06}. For a generalized measurement, this walk requires {\it feedback}---the weak measurement done at any stage in the process depends on the outcomes of measurements at earlier stages.  Continuous measurements with feedback have been shown to allow improvements in measurement efficiency \cite{Wiseman95}, so this is a subject of significant interest.  Some experimental systems can only be measured weakly.

For measurements with more than two outcomes, the measurement process corresponds to a random walk in a simplex \cite{Varbanov07}. However, these constructions assume that one can perform any weak measurement, which requires that one can do any weak joint unitary transformation on the system and ancilla.  In practice, this is not always the case, so it makes sense to also look at what kinds of measurements can be achieved with more restrictive families of weak measurements.

For a qubit system, a sufficient approach to decomposing any two-outcome measurements into a sequence of weak measurements is by introducing a fixed Hamiltonian (hence a fixed unitary) between the system and the successive ancilla probes (which are also assumed to be qubits).  In this approach, the measurement is controlled by the choice of the initial state and final projective measurement on the ancilla, with the feedback loop updating these choices at each step \cite{Florjanczyk14}. However, this approach is not sufficient to give any two-outcome measurement on a higher-dimensional system.  It can only produce final measurement operators with no more than two singular values.

A more powerful model \cite{Florjanczyk15} again characterizes the set of generalized measurements that can be decomposed into a sequence of weak measurements based on interactions between a stream of qubit probes and the system, but now the interaction Hamiltonian between them can itself be varied based on the feedback.  The work in \cite{Florjanczyk15} assumes that the interaction Hamiltonian is a linear combination of some fixed set of Hamiltonian terms.  The achievable measurements in this model have measurement operators from a finite dimensional Jordan algebra within the span of the Hamiltonian terms.  This approach generalizes to system states of any dimension.

This measurement model is more powerful, but the level of control needed for such joint Hamiltonians may be beyond the capability of many practical devices. Moreover, for some quantum systems, the natural measurement processes are {\it destructive}, as in photon counting measurements where each photon is destroyed when it is detected, and the measurement process ends when all the photons are consumed. Is it still possible to decompose a set of generalized measurements into a sequence of weak measurements when the measurements are destructive?

In this paper, we gain insight by constructing a simple qubit model of such a destructive weak measurement process, which uses a fixed weak swap unitary between the system and ancilla qubits followed by a projective measurement on the ancilla.  The initial state of the ancilla is always $|0\rangle \langle 0|$, and at long times the system qubit must inevitably be left in the state $|0\rangle\langle0|$ as well.  One can think of this weak measurement process as a toy model of a photon counting device, because the state $|0\rangle\langle 0|$, analogous to the vacuum state, is constantly leaking into the system.  But in this model we allow ourselves the freedom to do any projective measurement on the ancilla, which can depend on the outcomes of earlier measurements.  After measurement, the ancilla is consumed, and the next step of the procedure uses a fresh ancilla.

\subsection{The lossy qubit model}

In our model of a lossy measurement device, our system is a single qubit in an initial state $|\psi\rangle$, which interacts successively with a stream of ancilla qubits all prepared in the initial state $|0\rangle$. The interaction is a {\it weak swap} between the system qubit and the ancilla qubit, followed by a projective measurement of the ancilla. The weak swap is given by 
\begin{equation}
U=  I\cos \phi -i  S\sin \phi, 
\end{equation}
where $I$ is the identity and $S$ is the swap operator that exchanges the states of the system and the ancilla:
\begin{equation}
S|\psi\rangle \otimes |0\rangle= |0\rangle \otimes |\psi\rangle.
\end{equation}
The parameter $\phi$ controls the strength of the measurement, and is assumed to be small. We then do a projective measurement on the ancilla. If the projectors on the ancilla are $E_{k}=|e_{k}\rangle \langle e_{k}|$, where $\sum_{k}E_{k}=I$, the net effect will be a weak measurement on the state:
\begin{align}
|\psi\rangle \langle \psi| \to &
  \frac{\Tr_{\rm anc}\Big[\big( I \otimes E_{k}\big) U \big(|\psi\rangle \langle \psi| \otimes |0\rangle \langle0|\big) U^{\dagger} \big(I \otimes E_{k}\big) \Big]}{\Tr\Big[\big( I \otimes E_{k}\big) U \big(|\psi\rangle \langle \psi| \otimes |0\rangle \langle0|\big) U^{\dagger} \big(I \otimes E_{k}\big) \Big]}  \nonumber \\
& = \frac{M_{k} |\psi\rangle \langle \psi| M^{\dagger}_{k}}{\langle \psi |M^{\dagger}_{k}M_{k}| \psi \rangle} , 
\end{align}
where the operators
\begin{equation}
M_{k} = \langle e_{k} | 0\rangle \cos \phi \, I - i \sin\phi \, |0\rangle\langle e_{k}|
\label{eq:measOp}
\end{equation}
are the effective measurement operators on the state, and $\sum_{k}M^{\dagger}_{k}M_{k}=I$. After each step, the ancilla qubit is discarded and replaced by a fresh qubit in the state $|0\rangle$.

\section{Projective measurements of the system}

By choosing different projective measurements on the ancilla, one can perform a series of weak measurements on the system such that the whole set of measurements corresponds to a projective measurement. For an arbitrary two-outcome projective measurement with projectors $|b_0\rangle \langle b_0|$ and $|b_1\rangle \langle b_1|$ on the system, with initial state $|\psi\rangle$, one can find a series of projectors $E_{k_{i}}$ on the ancilla, where $k_{i}=0,1$, such that the product of the whole set of measurement operators $M_{k_1 \cdots k_{N}}\cdots M_{k_{1}}$ acts as the desired projective measurement, followed by a unitary rotation that moves the basis $\{|b_0\rangle , |b_1 \rangle\}$ to $\{|0\rangle, |1\rangle\}$. (The subscript $k_1 \cdots k_i$ indicates that the $i$th measurement operator depends on the whole previous outcomes.) The procedure is as follows.

\subsection{The construction and its properties}

We will use the polar decomposition for the measurement operators:
\[
M_{k}=U_k\sqrt{M^\dagger_k M_k} .
\]
We choose projectors $E_k=|e_k\rangle \langle e_k|$ on the ancilla such that 
\begin{equation}
M_{k}=U_{k} \Big(\sqrt{\lambda_{0,k}}|b_{0}\rangle\langle b_{0}|+\sqrt{\lambda_{1,k}}|b_{1}\rangle\langle b_{1}| \Big), \ \ \ k=0,1,
\label{eq:polarDecomp}
\end{equation}
where $\lambda_{0,k}$ and $\lambda_{1,k}$ are the eigenvalues for the POVM elements:
\begin{equation}
M^{\dagger}_{k}M_{k}=\lambda_{0,k}|b_{0}\rangle\langle b_{0}|+\lambda_{1,k}|b_{1}\rangle\langle b_{1}|.
\label{eq:povmDiagonal} 
\end{equation}
We can always find projectors $E_k$ on the ancilla such that Eq.~(\ref{eq:povmDiagonal}) is satisfied for any given measurement basis $\{|b_0\rangle, |b_1\rangle\}$.
If the measurement basis is $\{|b_0\rangle =a |0\rangle + b e^{i \chi}|1\rangle, |b_1\rangle =b |0\rangle -a e^{i \chi}|1\rangle\}$, where $a , b \geq 0$, $a^2+b^2=1$ and $\chi$ is real, then the basis $\{|e_0\rangle,|e_1\rangle\}$ for the projectors $E_{0}, E_1$ on the ancilla should be chosen as
\begin{equation}
|e_{0}\rangle,|e_1\rangle= \begin{cases} \sqrt{\frac{1\pm \delta}{2}}|0\rangle \pm \sqrt{\frac{1\mp\delta}{2}} e^{i \theta}|1\rangle & \mbox{if} \ a  \geq b, \\
        \sqrt{\frac{1\mp\delta}{2}}|0\rangle \pm \sqrt{\frac{1\pm\delta}{2}} e^{i \theta}|1\rangle & \mbox{if} \ a  \leq b, \end{cases}
\end{equation}
where
\[
\delta=\sqrt{\frac{1-4a^2b^2}{1-4a^2b^2 \cos^2\phi}}
\]
and $\theta=\chi+\phi-\pi/2$.  The eigenvalues for the POVM elements in Eq.~(\ref{eq:povmDiagonal}) will be 
\begin{eqnarray}
\lambda_{0,0} = \lambda_{1,1} &=& \frac{1}{2}\Big(1+\frac{\sqrt{1-4a^2 b^2}\cos^2\phi + \sin^2\phi}{\sqrt{1-4a^2 b^2\cos^2\phi}}\Big) , \nonumber\\ 
\lambda_{0,1} = \lambda_{1,0} &=& \frac{1}{2}\Big(1+\frac{\sqrt{1-4a^2 b^2}\cos^2\phi - \sin^2\phi}{\sqrt{1-4a^2 b^2\cos^2\phi}}\Big) .
\label{eq:eigenvalues}
\end{eqnarray}
The resulting unitary in Eq.~(\ref{eq:polarDecomp}) then becomes 
\begin{equation}
U_k = M_k \Big( \frac{1}{\sqrt{\lambda_{0,k}}}|b_0\rangle\langle b_0| + \frac{1}{\sqrt{\lambda_{1,k}}}|b_1\rangle \langle b_1|\Big),
\end{equation}
where  $M_k$ is the operator in Eq.~(\ref{eq:measOp}), when $\lambda_{0,1,k} \neq 0$, which is true for $\{|b_{0}\rangle,|b_{1}\rangle\} \neq \{|0\rangle,|1\rangle\}$. If $|b_{0}\rangle = |0\rangle$ and $|b_1\rangle = |1\rangle$, then the unitaries will be $U_0 =    \begin{pmatrix} % or pmatrix or bmatrix or Bmatrix or ...
      e^{-i \phi} & 0 \\
      0 & 1 \\
   \end{pmatrix}$ and $U_1 = -i\sigma_x$; If $|b_{0}\rangle = |1\rangle$ and $|b_1\rangle = |0\rangle$, then the unitaries will be $U_0 = -i\sigma_x$ and $U_1 = \begin{pmatrix} % or pmatrix or bmatrix or Bmatrix or ...
      e^{-i \phi} & 0 \\
      0 & 1 \\
   \end{pmatrix}$.

By this construction, the first weak measurement for a projective measurement $|b_0\rangle\langle b_0|$ and $|b_1\rangle\langle b_1|$ on a state $|\psi\rangle= \alpha |b_0\rangle +\beta |b_1\rangle$ would be 
\begin{equation}
M_{k_1}=U_{k_1} (\sqrt{\lambda_{0,k_1}}|b_{0}\rangle\langle b_{0}|+\sqrt{\lambda_{1,k_1}}|b_{1}\rangle\langle b_{1}|), \ \ k_1=0,1.
\end{equation}
Depending on the outcome $k_1$ from the first measurement, the basis vectors $|b_0\rangle$ and $|b_1 \rangle$ will be moved to $|b^{k_1}_0\rangle= U_{k_1}|b_0\rangle$ and $|b^{k_1}_1\rangle= U_{k_1}|b_1\rangle$. The state after the first measurement will take the form
\begin{align}
|\psi_{k_1}\rangle &= M_{k_1}|\psi\rangle/\sqrt{p_{k_1}} \nonumber\\
 &= \frac{1}{\sqrt{p_{k_1}}}\Big(\alpha \sqrt{\lambda_{0,k_1}} |b^{k_1}_{0}\rangle +\beta \sqrt{\lambda_{1,k_1}} |b^{k_1}_{1}\rangle\Big),
\end{align}
where  $p_{k_1}$ is the probability of outcome $k_1$:
\begin{equation}
p_{k_1}= \langle \psi| M^{\dagger}_{k_1}M_{k_1}|\psi\rangle = |\alpha|^2\lambda_{0,k_1} +|\beta|^2 \lambda_{1,k_1} .
\end{equation}
Because the measurement basis changes from $|b_{0,1}\rangle$ to $|b^{k_1}_{0,1}\rangle$, the second set of measurement operators are
\begin{equation}
M_{k_1 k_2}=U_{k_1 k_2} \Big(\sqrt{\lambda_{0,k_1k_2}}|b^{k_1}_{0}\rangle\langle b^{k_1}_{0}|+\sqrt{\lambda_{1,k_1k_2}}|b^{k_1}_{1}\rangle\langle b^{k_1}_{1}| \Big),   
\end{equation} 
$k_{1,2}=0,1$, where $\lambda_{0,k_1k_2}$ and $\lambda_{1,k_1k_2}$ are the two eigenvalues of the POVM element $M^{\dagger}_{k_1 k_2}M_{k_1 k_2}$ given that the first outcome is $k_1$. After the second measurement, the state becomes
\begin{align}
|\psi_{k_1 k_2}\rangle&= \frac{1}{\sqrt{p_{k_2|k_1}}} M_{k_1 k_2} |\psi_{k_1}\rangle= \frac{1}{\sqrt{p_{k_2|k_1}p_{k_1}}}M_{k_1 k_2}M_{k_1}|\psi\rangle \nonumber \\
&= \frac{1}{\sqrt{p_{k_1 k_2}}} \Big(\alpha \sqrt{\lambda_{0,k_1}\lambda_{0,k_1 k_2}}|b^{k_1 k_2}_{0}\rangle \nonumber \\
& \ \ \ \ \ \ \ \ \ \ \ \ \ +\beta \sqrt{\lambda_{1,k_1}\lambda_{1,k_1 k_2}} |b^{k_1 k_2}_{1}\rangle \Big),
\end{align}
where $p_{k_1 k_2}$ is the probability of obtaining the outcomes $k_1$ and $k_2$, and $|b^{k_1 k_2}_{0,1}\rangle = U_{k_1 k_2} U_{k_1} |b_{0,1}\rangle$ is the measurement basis after the first two measurements. By continuing the process $N$ times, the state becomes
\begin{align}
 &|\psi_{k_1 \cdots k_N}\rangle = \frac{1}{\sqrt{p_{k_1 \cdots k_N}}} M_{k_1 \cdots k_N} \cdots M_{k_1} |\psi \rangle \nonumber \\
& = \frac{1}{\sqrt{p_{k_1 \cdots k_N}}}\Big( \sqrt{\lambda_{0,k_1}\cdots \lambda_{0,k_1 \cdots k_N}}|b^{k_1 \cdots k_N}_{0}\rangle \langle b_0| \nonumber \\
&\ \ \ \ \ \ \ \ \ \ \ \ \ \ \ \ + \sqrt{\lambda_{1,k_1} \cdots \lambda_{1,k_1 \cdots k_N}}|b^{k_1 \cdots k_N}_{1}\rangle \langle b_1| \Big) |\psi\rangle \nonumber \\
&= \frac{1}{\sqrt{p_{k_1 \cdots k_N}}}\Big(\alpha \sqrt{\lambda_{0,k_1}\cdots \lambda_{0,k_1 \cdots k_N}}|b^{k_1 \cdots k_N}_{0}\rangle \nonumber \\
&\ \ \ \ \ \ \ \ \ \ \ \ \ \ \ \ +\beta \sqrt{\lambda_{1,k_1} \cdots \lambda_{1,k_1\cdots k_N}}|b^{k_1 \cdots k_N}_{1}\rangle \Big),
\label{eq:stateNstep}
\end{align}
where $p_{k_1 \cdots k_N}$ is the probability of getting the string of outcomes $k_1 \cdots k_N$:
\begin{eqnarray}
p_{k_1 \cdots k_N} = && |\alpha|^2 \lambda_{0,k_1}\cdots \lambda_{0,k_1\cdots k_N} \nonumber\\
&& + |\beta|^2 \lambda_{1,k_1}\cdots \lambda_{1,k_1\cdots k_N}.
\label{eq:stringProb}
\end{eqnarray}

As the number $N$ increases, the measurement basis $|b^{k_1 \cdots k_N}_{0,1}\rangle$ approaches to $|0\rangle ,|1\rangle$. This can be proved by looking at the form of the sequence of measurements $M_{k_1 \cdots k_N} \cdots M_{k_1} $. Each measurement operator $M_{k_1 \cdots k_i}$ is a $2\times2$ matrix of the form
\begin{equation}
M_{k_1 \cdots k_i}=\Bigg(\begin{matrix}
\langle e_{k_i}|0\rangle e^{-i\phi} & -i \langle e_{k_i}|1\rangle \sin\phi \\
0 & \langle e_{k_i}|0\rangle \cos\phi
\end{matrix}\Bigg).
\end{equation}

Without loss of generality, one can choose $\langle e_{k_i}|0\rangle $ to be a positive real number with the projector $|e_{k_i}\rangle\langle e_{k_i}|$ on the ancilla. Furthermore, if the initial projective measurement $|b_{0,1}\rangle$ is not $|0\rangle,|1\rangle$ then $\langle e_{k_i}|0\rangle$ will not be zero at any point in the process. This can be shown as follows:  assume that one starts from a projective measurement $\{|b_{0}\rangle,|b_1\rangle\} \neq \{|0\rangle,|1\rangle\}$ (that is, neither $|b_0\rangle=|0\rangle$ nor $|b_0\rangle=|1\rangle$).  Now suppose that at the $i$th measurement one gets $\langle e_{k_i}|0\rangle=0$. This implies that the measurement operator for the $i$th measurement is $M_{k_i}=-i \sin\phi |0\rangle\langle1|$, which means that the $(i-1)$th measurement is already in the $\{|0\rangle,|1\rangle\}$ basis. One can easily show that the $(i-2)$th measurement then must also be in the $\{|0\rangle,|1\rangle\}$ basis, and so on. Hence, one can deduce that for the first measurement $\{|b_{0}\rangle,|b_1\rangle\} = \{|0\rangle,|1\rangle\}$. This contradicts the assumption, and therefore no $\langle e_{k_i}|0\rangle$ can be zero throughout the whole process. This also shows that the measurement basis will never be $\{|0\rangle,|1\rangle\}$ at any point in the process if one starts from $\{|b_{0}\rangle,|b_1\rangle\} \neq \{|0\rangle,|1\rangle\}$, though the measurement basis will approach $\{|0\rangle,|1\rangle\}$ asymptotically.

If one performs such a sequence of measurements, the product of the measurement operators becomes
\begin{equation}
M_{k_1 \cdots k_N} \cdots M_{k_1}=\begin{pmatrix}
1 & x \\
0& \epsilon
\end{pmatrix} \displaystyle\prod_{i=1}^{N}\langle e_{k_i}|0\rangle,
\end{equation}
where 
\begin{equation}
x=-i e^{i N\phi}\sin\phi \Big(\displaystyle\sum^{N}_{j=1} \frac{\langle e_{k_j}|1\rangle}{\langle e_{k_j}|0\rangle} e^{-i \phi (N-j)} \cos^{j-1}\phi\Big)
\end{equation}
and 
\begin{equation}
\epsilon =e^{i N\phi}\cos^{N}{\phi}.
\end{equation}

From the second line in Eq.~(\ref{eq:stateNstep}), the measurement basis $|b^{k_1\cdots k_N}_{0,1}\rangle$ after $N$ measurements is the eigenbasis of the matrix
\begin{align}
M_{k_1 \cdots k_N} \cdots & M_{k_1}M^{\dagger}_{k_1} \cdots M^{\dagger}_{k_1 \cdots k_N}  \nonumber \\
&= \lambda_{0,k_1}\cdots \lambda_{0,k_1\cdots k_N}|b^{k_1 \cdots k_N}_{0}\rangle \langle b^{k_1 \cdots k_N}_{0}| \nonumber \\
&\ + \lambda_{1,k_1} \cdots \lambda_{1,k_1\cdots k_N}|b^{k_1 \cdots k_N}_{1}\rangle \langle b^{k_1 \cdots k_N}_{1}| \nonumber \\
& \propto \begin{pmatrix}
1 & x \\
0 & \epsilon
\end{pmatrix} 
\begin{pmatrix}
1 & x \\
0& \epsilon 
\end{pmatrix}^{\dagger}=
\begin{pmatrix}
1+|x|^2 & x \epsilon^*\\
 x^* \epsilon& |\epsilon |^2
\end{pmatrix}.
\label{eq:asymptoticProduct}
\end{align}
Let $|v_+\rangle, |v_-\rangle$ be the two eigenvectors of the matrix in Eq.~(\ref{eq:asymptoticProduct}), and define $|v_+\rangle$ to be the eigenvector that is closer to $|0\rangle$. The squared magnitude of the inner product of $|v_+\rangle$ with $|0\rangle$ is
\begin{align}
&|\langle v_+|0\rangle|^2 = \nonumber \\
&\frac{\big( 1+ |x|^2- |\epsilon|^2 + \sqrt{(1+|x|^2+|\epsilon|^2)^2-4 |\epsilon|^2} \big)^2}{\big( 1+ |x|^2- |\epsilon|^2 + \sqrt{(1+|x|^2+|\epsilon|^2)^2-4 |\epsilon|^2} \big)^2 + 4|x|^2 |\epsilon|^2}.
\end{align}

In the large $N$ limit, $|\epsilon|^2$ goes to 0 like $\cos^{2N}\phi$, and $|\langle v_+|0\rangle|^2$ approaches 1 no matter what value $|x|^2$ takes. This can be shown as follows. If $|x|^2$ approaches 0, then $|\langle v_+|0\rangle|^2\approx1 -|x|^2|\epsilon|^2$. If $|x|^2$ approaches infinity, then $|\langle v_+|0\rangle|^2\approx1 -(1/|x|^2)|\epsilon|^2$. If $|x|^2$ approaches a finite number, then $|\langle v_+|0\rangle|^2\approx 1- (|x|^2/(1+|x|^2)^2) |\epsilon|^2 $. By the orthogonality of the eigenbasis, $|\langle v_-|0\rangle|^2$ therefore approaches 0. Hence, the measurement basis $|b^{k_1\cdots k_N}_{0,1}\rangle$, which is equivalent to $|v_{\pm}\rangle$, must approach $\{|0\rangle ,|1\rangle\}$.

By looking at the quantity $|\langle b^{k_1\cdots k_N}_{0}|0\rangle|^2$, if $|\langle b^{k_1\cdots k_N}_{0}|0\rangle|^2 \to 1$ we conclude that the outcome is $|b_0\rangle$, while if $|\langle b^{k_1\cdots k_N}_{0}|0\rangle|^2 \to 0$ we conclude that the outcome is $|b_1\rangle$. We will now show that the probability of concluding the outcome to be $|b_{0}\rangle$ (or $|b_1\rangle$) approaches $|\alpha|^2$ (or $|\beta|^2$) in the large $N$ limit.

\subsection{The case of $\{|0\rangle,|1\rangle\}$ as a boundary condition}

Let us first consider the case where the projective measurement to be performed on the state is $|b_{0}\rangle=|0\rangle$, $|b_{1}\rangle=|1\rangle$. By choosing the projectors $|e_{0}\rangle =|0\rangle ,|e_1 \rangle= |1\rangle$, the first measurement operators on the state will be 
\begin{equation}
M_0=e^{-i \phi}|0\rangle\langle 0|+ \cos\phi |1\rangle\langle 1|,
\end{equation}
and
\begin{equation}
M_1=-i\sin\phi |0\rangle\langle1|.
 \end{equation}

If one gets outcome 0, then the new measurement basis remains the same: $|b^{0}_{0}\rangle = |0\rangle$ and $|b^{0}_{1}\rangle = |1\rangle$.  The initial state $|\psi\rangle=\alpha |0\rangle +\beta |1\rangle$ becomes
 \begin{equation}
 |\psi_0\rangle=\frac{1}{\sqrt{p_0}}\big(\alpha e^{-i\phi} |0\rangle + \beta \cos\phi |1\rangle\big), 
 \end{equation}
where $p_{0}=|\alpha|^2+|\beta|^2 \cos^2{\phi}$ is the probability of outcome 0.  One then repeats the same process for the second measurement. If at any point one gets the outcome 1, then the new measurement basis is flipped, $|b^{1}_{0}\rangle=|1\rangle$ and $|b^{1}_{1}\rangle=|0\rangle$, and the system state goes to $|0\rangle$. The process can be terminated at that point, because any subsequent measurements will get $|0\rangle$ on the ancilla with probability 1.  Then we conclude the outcome is 1. If one never gets the result $|1\rangle$ on the ancilla, then we conclude the outcome is 0. If one repeats the measurements $N$ times then the probability of concluding that the outcome is 0 (namely, never getting $|1\rangle$ on the ancilla), is $|\alpha|^2+|\beta|^2 \cos^{2N}{\phi}$, and the probability of concluding that the outcome is 1 is $1-(|\alpha|^2+|\beta|^2 \cos^{2N}{\phi})=|\beta|^2(1 - \cos^{2N}{\phi})$. In the large $N$ limit, the probability of concluding that the outcome is $|0\rangle$ or $|1\rangle$ approaches $|\alpha|^2$ or $|\beta|^2$, respectively. This accomplishes a projective measurement $\{|0\rangle, |1\rangle\}$ on the state, and this special case sets a boundary condition for the model.

\subsection{Outcome probabilities in the general case}

We now show that the probabilities of the two outcomes match the usual Born rule for a general projective measurement in the basis $\{|b_0\rangle, |b_1\rangle \}$.  For the initial state $|\psi\rangle=\alpha |b_0\rangle +\beta |b_1\rangle$, we define the function $P\left(|b_{0,1}\rangle,\alpha\right)$ to be the probability of concluding that the outcome is $|b_0\rangle$.  This is equivalent to the condition that $|\langle b^{k_1\cdots k_N}_{0}|0\rangle|^2 \to 1$ in the large $N$ limit. $P$ does not depend on the relative phase between $\alpha$ and $\beta$ or the relative phase $\chi$ in basis $|b_{0,1}\rangle$, because only the magnitude of $\alpha$ and $a$ enter the expression for the probabilities for outcomes as shown in Eq. (\ref{eq:eigenvalues}) and Eq. (\ref{eq:stringProb}). From probability theory we have the following equation:
\begin{eqnarray}
P\left(|b_{0,1}\rangle,\alpha\right) &=& p_0P\left(|b^{0}_{0,1}\rangle, \frac{\alpha \sqrt{\lambda_{0,0}}}{\sqrt{p_0}}\right) \nonumber\\
&& + p_1 P\left(|b^{1}_{0,1}\rangle, \frac{\alpha \sqrt{\lambda_{0,1}}}{\sqrt{p_1}}\right),
\end{eqnarray}
where $p_{0}$ is the probability of getting outcome $0$ from the first measurement and $P\left(|b^{0}_{0,1}\rangle, \alpha \sqrt{\lambda_{0,0}/p_0}\right)$ is the probability to conclude the outcome is $|b^0_0\rangle$ starting from the new state and measurement basis after the first measurement. If we iterate this expression for $N$ times, it becomes
\begin{align}
& P(|b_{0,1}\rangle,\alpha)= p_0P\Big(|b^{0}_{0,1}\rangle, \frac{\alpha \sqrt{\lambda_{0,0}}}{\sqrt{p_0}}\Big)+p_1 P\Big(|b^{1}_{0,1}\rangle, \frac{\alpha \sqrt{\lambda_{0,1}}}{\sqrt{p_1}}\Big)  \nonumber \\
 &\mathsmaller{=p_0\Bigg[ p_{0|0}P\Big(|b^{00}_{0,1}\rangle, \frac{\alpha \sqrt{\lambda_{0,0}\lambda_{0,00}}}{\sqrt{p_{00}}} \Big) + p_{1|0}P\Big(|b^{01}_{0,1}\rangle, \frac{\alpha \sqrt{\lambda_{0,0}\lambda_{0,01}}}{\sqrt{p_{01}}} \Big) \Bigg]}  \nonumber \\
   &\mathsmaller{+ p_1\Bigg[ p_{0|1}P\Big(|b^{10}_{0,1}\rangle, \frac{\alpha \sqrt{\lambda_{0,1}\lambda_{0,10}}}{\sqrt{p_{10}}} \Big) + p_{1|1}P\Big(|b^{11}_{0,1}\rangle, \frac{\alpha \sqrt{\lambda_{0,1}\lambda_{0,11}}}{\sqrt{p_{11}}} \Big) \Bigg] } \nonumber \\
& \ \ \ \ \ \ \ \ \ \ \ \ \ \ \ \ \ \ \ \ \ \ \ \ \ \ \ \ \ \ \ \ \ \ \  \vdots    \nonumber \\
 & = \displaystyle\sum_{k_1\cdots k_N=0,1} p_{k_1\cdots k_N} P\Big( |b^{k_1 \cdots k_N}_{0,1}\rangle, \frac{\alpha \sqrt{\lambda_{0,k_1} \cdots \lambda_{0,k_1\cdots k_N}}}{\sqrt{p_{k_1\cdots k_N}}} \Big).
\label{eq:probFuncN}
\end{align}
From the special case above, we have $P(|0,1\rangle,\alpha)=|\alpha|^2+|\beta|^2 \cos^{2N}\phi$, and
\begin{equation}
|P(|0,1\rangle,\alpha)-|\alpha|^2|\leq \cos^{2N}\phi.
\label{eq:probBound}
\end{equation}
By the continuity of the function $P(|b_{0,1}\rangle,\alpha)$ at $|b_{0,1}\rangle=|0,1\rangle$, there exists a quantity $\delta>0$ such that
\begin{equation}
|P(|b_{0,1}\rangle,\alpha)-P(|0,1\rangle,\alpha)| \leq  \delta,
\label{eq:continuity}
\end{equation}
and as $|b_{0,1}\rangle \to |0,1\rangle$, $\delta \to 0$. Now we can evaluate the function $P$ in Eq.~(\ref{eq:probFuncN}). From the result given earlier, the measurement basis $|b^{k_1\cdots k_N}_{0,1}\rangle$ approaches $|0,1\rangle$ for any outcomes $k_1\cdots k_N$ in the large $N$ limit. By combining Eq.~(\ref{eq:probBound}) and Eq.~(\ref{eq:continuity}), the function $P$ in Eq.~(\ref{eq:probFuncN}) has the following bound:
\begin{align}
& \mathsmaller{\Bigg| P\Big( |b^{k_1 \cdots k_N}_{0,1}\rangle, \frac{\alpha \sqrt{\lambda_{0,k_1} \cdots \lambda_{0,k_1\cdots k_N}}}{\sqrt{p_{k_1\cdots k_N}}} \Big)-\frac{|\alpha|^2 \lambda_{0,k_1}\cdots \lambda_{0,k_1 \cdots k_N}}{p_{k_1\cdots k_N}} \Bigg|} \nonumber \\
& \leq \delta(k_1 \cdots k_N) +\cos^{2N}\phi,
\label{eq:bound}
\end{align}
where $\delta(k_1\cdots k_N)$ is the quantity in Eq.~(\ref{eq:continuity}) for a certain combination of outcomes $k_1 \cdots k_N$. With the expression (\ref{eq:bound}), Eq.~(\ref{eq:probFuncN}) becomes 
\begin{align}
 &P(|b_{0,1}\rangle,\alpha) \nonumber \\
 &=\displaystyle\sum_{k_1\cdots k_N=0,1} p_{k_1\cdots k_N} P\Big( |b^{k_1 \cdots k_N}_{0,1}\rangle, \frac{\alpha \sqrt{\lambda_{0,k_1} \cdots \lambda_{0, k_1\cdots k_N}}}{\sqrt{p_{k_1\cdots k_N}}} \Big) \nonumber \\
 &\leq \displaystyle\sum_{k_1\cdots k_N=0,1} \Bigg( |\alpha|^2\lambda_{0,k_1} \cdots \lambda_{0,k_1 \cdots  k_N} \nonumber \\
 & \ \ \ \ \ \ \ \ \ \ \ \ \ \ + p_{k_1 \cdots k_N} \Big[ \max\delta(k_1 \cdots k_N) + \cos^{2N}\phi\Big]\Bigg) \nonumber \\
 &= |\alpha|^2 + \max\delta(k_1 \cdots k_N) + \cos^{2N}\phi,
\label{eq:bound2}
\end{align}
where $\max\delta(k_1 \cdots k_N)=\delta(k^{*}_{1} \cdots k^{*}_{N}) $ is the maximum variation, which occurs for a certain string of outcomes $k^{*}_{1} \cdots k^{*}_{N}$. The last equality in (\ref{eq:bound2}) is due to Eq.~(\ref{eq:stringProb}):
\begin{align}
&\displaystyle\sum_{k_1 \cdots k_N=0,1} p_{k_1\cdots k_N} =1 \nonumber \\
& =|\alpha|^2 \displaystyle\sum_{k_1 \cdots k_N=0,1} \lambda_{0,k_1}\cdots \lambda_{0,k_1 \cdots k_N} \nonumber \\
& \ \ +|\beta|^2 \displaystyle\sum_{k_1 \cdots k_N=0,1}\lambda_{1,k_1}\cdots \lambda_{1,k_1 \cdots k_N}
\end{align}
is satisfied for all $\alpha$ and $\beta$. Hence,
\begin{align}
 &\displaystyle\sum_{k_1 \cdots k_N=0,1} \lambda_{0,k_1}\cdots \lambda_{0,k_1\cdots k_N}=\displaystyle\sum_{k_1 \cdots k_N=0,1} \lambda_{1,k_1}\cdots \lambda_{1,k_1 \cdots k_N} \nonumber \\
 &\ \ \ \ =1
\end{align}
All $\delta(k_1 \cdots k_N)$ and $\cos^{2N}\phi$ approach zero in the large $N$ limit, and therefore $P(|b_{0,1}\rangle,\alpha)$ approaches $|\alpha|^2$. This shows that by using the lossy device, one can perform any projective measurement on a state.

\section{POVMs with commuting POVM elements}

We have shown that the destructive model can implement an arbitrary projective measurement. There is a simple way to perform a more general POVM with any set of mutually commuting POVM elements, by introducing a random number generator together with the lossy device and doing an extra classical postprocessing step.  Because commuting POVM elements share the same eigenvectors, they can be expanded in the same eigenbasis.  We perform a projective measurement in this eigenbasis, followed by a probabilistic output step.

For simplicity, consider a POVM with three commuting POVM elements:
\begin{align}
&\mathcal{E}_1= a |b_0\rangle \langle b_0| + b |b_1\rangle \langle b_1|, \nonumber \\
&\mathcal{E}_2= c |b_0\rangle \langle b_0| + d |b_1\rangle \langle b_1|, \nonumber \\
&\mathcal{E}_3= (1-a-c) |b_0\rangle \langle b_0| + (1-b-d)|b_1\rangle \langle b_1|,
\end{align}
where $1\geq a,b,c,d \geq 0$ and $1 \geq a+c, 1\geq b+d$. The eigenbasis is $\{|b_0\rangle,|b_1\rangle\}$. For an initial state $|\psi\rangle= \alpha |b_0\rangle + \beta |b_1\rangle$, the probabilities of outcomes 1, 2, and 3 are
\begin{align}
&\mathcal{P}_1= a |\alpha|^2 + b |\beta|^2 , \nonumber \\
&\mathcal{P}_2= c |\alpha|^2  + d |\beta|^2 , \nonumber \\
&\mathcal{P}_3= (1-a-c)|\alpha|^2  + (1-b-d)|\beta|^2.
\label{eq:3outcomeProbs}
\end{align}
One first performs a $|b_0\rangle,|b_1\rangle$ projective measurement using the lossy model in the previous section.  One then generates the output using a random number generator conditioned on the outcome of the projective measurement.  For projective measurement outcome $|b_0\rangle$, we output 1, 2, or 3 with probabilities $a$, $c$, and $1-a-c$, respectively; for projective measurement outcome $|b_1\rangle$, we output 1, 2 or 3 with the probabilities $b$, $d$, or $1-b-d$, respectively. Then the unconditioned probabilities of getting outcomes 1, 2, and 3 will be the same as Eq.~(\ref{eq:3outcomeProbs}). The outcome of the projective measurement and the values of the random numbers are discarded. One can think of the lossy projective measurement process and the random number generator together as parts of a single device:  the outcome from the projective measurement is input into the random number generator, which gives the final outcome 1, 2, or 3. One can easily generalize this approach into a POVM with an arbitrary number of commuting POVM elements.

\section{Conclusion}\label{conclusion}

It has been previously shown that any generalized measurement can be performed as a sequence of weak measurements if one has the ability to perform any weak measurement.  If only a limited family of weak measurements is possible, however, the class of generalized measurements that can be done may also be limited.  This has been explored for different plausible families of weak measurements.  In this paper we have significantly generalized that earlier work by having the weak measurements be {\it destructive}:  at long times, the system always goes to a fixed state $|0\rangle$.  The measurements that can be done in this case must obviously also be destructive:  POVMs, rather than ideal generalized measurements.  We considered a simple model of this case where both the system and the probes are qubits, which interact by a weak swap unitary, and where the initial probe state is always $|0\rangle$.  Surprisingly, it is still possible to do any projective measurement, in any basis, even in this highly restrictive model.  We gave a constructive procedure for doing such a measurement by a sequence of lossy weak measurements, and proved that this procedure gives the same outcomes as a strong projective measurement, with the same probabilities.  We also showed that this implies the ability to do any POVM with mutually commuting POVM elements.

The success of this simple model raises two very significant questions.  First, is it possible to generalize this construction to allow general POVMs, including those with noncommuting elements?  We believe the answer to this question is yes, and will publish our method in forthcoming work.  It requires a significant generalization of the current construction.

Second, do these results generalize to higher-dimensional systems?  It may well be that techniques that are sufficient to allow any POVM on a qubit may only permit a restricted class of measurements on higher-dimensional systems; similar restrictions have been seen in other cases \cite{Florjanczyk14}.  The answer may also depend on what constitutes a physically realistic generalization to higher-dimensional systems.  This is a large and complicated question, and our work here is ongoing.  The problem of continuous measurements remains interesting and rich, and we hope that these methods may in time open up the possibility of new types of practical experimental measurements.

\section*{Acknowledgments}

YHC and TAB acknowledge useful conversations with Christopher Cantwell, Ivan Deutsch, Shengshi Pang,  Christopher Sutherland, and Howard Wiseman.  This work was funded in part by  the ARO MURI under Grant No. W911NF-11-1-0268, NSF Grant No. CCF-1421078, NSF Grant No. QIS-1719778, and by an IBM Einstein Fellowship at the Institute for Advanced Study.

\bibliography{lossy_weak} 

\end{document}